\documentclass[conference]{IEEEtran}
\IEEEoverridecommandlockouts
\usepackage{cite}
\usepackage{amsmath,amssymb,amsfonts,amsthm}
\usepackage{algorithmic}
\usepackage{graphicx}
\usepackage{textcomp}
\usepackage{xcolor}
\usepackage{authblk}
\usepackage[a4paper, total={184mm,239mm}]{geometry}
\usepackage{pifont}
\usepackage[hidelinks]{hyperref}
\usepackage[table]{xcolor} 
\usepackage{braket} 
\usepackage{cleveref}
\usepackage{booktabs}
\usepackage{quantikz}
\usepackage{subfigure}
\usepackage[linesnumbered,ruled,vlined]{algorithm2e}%

\def\BibTeX{{\rm B\kern-.05em{\sc i\kern-.025em b}\kern-.08em
    T\kern-.1667em\lower.7ex\hbox{E}\kern-.125emX}}
\begin{document}

\newtheorem{definition}{Definition}
\newtheorem{theorem}{Theorem}
\newtheorem{proposition}{Proposition}
\newtheorem{lemma}{Lemma}
\newtheorem{corollary}{Corollary}
\newtheorem{example}{Example}
\newtheorem{conjecture}{Conjecture}

\Crefname{figure}{Fig.}{Figs.}

\title{Optimal Compilation of Syndrome Extraction Circuits for General Quantum LDPC Codes}

\author[1, 2]{Kai Zhang}
\author[3]{Dingchao Gao}
\author[4]{Zhaohui Yang}
\author[1]{Runshi Zhou}
\author[2]{Fangming Liu}
\author[1]{Zhengfeng Ji}
\author[1, $\dagger$]{Jianxin Chen}

\affil[1]{Department of Computer Science and Technology, Tsinghua University}
\affil[2]{Pengcheng Laboratory}

\affil[3]{Key Laboratory of System Software (Chinese Academy of Sciences), }
\affil[ ]{Institute of Software, Chinese Academy of Sciences}

\affil[4]{Department of Electronic and Computer Engineering, The Hong Kong University of Science and Technology}

\affil[$\dagger$]{Corresponding author: chenjianxin@tsinghua.edu.cn}

\maketitle

\begin{abstract}
Quantum error correcting codes (QECC) are essential for constructing large-scale quantum computers that deliver faithful results. As strong competitors to the conventional surface code, quantum low-density parity-check (qLDPC) codes are emerging rapidly: they offer high encoding rates while maintaining reasonable physical-qubit connectivity requirements. Despite the existence of numerous code constructions, a notable gap persists between these designs---some of which remain purely theoretical---and their circuit-level deployment.

In this work, we propose Auto-Stabilizer-Check (ASC), a universal compilation framework that generates depth-optimal syndrome extraction circuits for arbitrary qLDPC codes. ASC leverages the sparsity of parity-check matrices and exploits the commutativity of X and Z stabilizer measurement subroutines to search for optimal compilation schemes. By iteratively invoking an SMT solver, ASC returns a depth-optimal solution if a satisfying assignment is found, and a near-optimal solution in cases of solver timeouts. Notably, ASC provides the first definitive answer to one of IBM's open problems: for all instances of bivariate bicycle (BB) code reported in their work, our compiler certifies that no depth-6 syndrome extraction circuit exists.

Furthermore, by integrating ASC with an end-to-end evaluation framework---one that assesses different compilation settings under a circuit-level noise model---ASC reduces circuit depth by approximately 50\% and achieves an average 7x-8x suppression of the logical error rate for general qLDPC codes, compared with as-soon-as-possible (ASAP) and coloration-based scheduling. ASC thus substantially reduces manual design overhead and demonstrates its strong potential to serve as a key component in accelerating hardware deployment of qLDPC codes.
\end{abstract}


\section{Introduction}
Quantum computing harnesses the unique principles of quantum mechanics and holds the potential to deliver significant speedups for various hard problems~\cite{shor1999polynomial,grover1996fast,peruzzo2014variational, gidney2025factor}. While there is no ``free lunch'' in quantum computer development---given the inherently noisy environment and the fragility of qubits---quantum error correction (QEC) enables reliable quantum computing by using redundant physical qubits for logical encoding~\cite{aharonov1997fault, kitaev1997quantum, Knill1998ResilientQC}, thereby enabling the high-fidelity storage and processing of quantum information~\cite{litinski2019game}.

Building a robust, fault-tolerant quantum computer depends not only on high-performance QECCs~\cite{Dennis_2002, hastings2021dynamically, bravyi2024high, Bravyi_2024_arxiv} and efficient decoding architectures~\cite{higgott2025sparse, bausch2024learning, wu2023fusion, Tan2022ScalableSD, muller2025improved, zhang2025learning}, but also on the effective design of syndrome extraction circuits,
which is a compiler optimization task serving as a crucial bridge between code abstraction and physical gate execution. While bespoke and experimentally validated~\cite{Acharya2022SuppressingQE, bravyi2024high, Bravyi_2024_arxiv}, hand-designed syndrome extraction circuits exist for specific codes---such as the depth-4 surface code or depth-7 BB code---these require substantial manual efforts and experience. Consequently, a compiler capable of generating optimal circuits for arbitrary qLDPC codes remains a critical missing piece in the field.

Previous compilers designed for NISQ are not tailored for QEC circuits. Compilers such as Qiskit~\cite{javadi2024quantum} and TKet~\cite{sivarajah2020t} utilize peephole optimization based on gate cancellations and commutativity rules within local subcircuits and schedule gate sequences in the as-soon-as-possible (ASAP) approach. However, they cannot lead to non-trivial optimization effects for syndrome extraction circuits, as they only consider local commutativity optimization opportunities. Tremblay et al.~\cite{tremblay2022constant} observed the commutativity within the same-type stabilizer checks and introduced a coloration-based method to schedule X and Z stabilizer checks of the CSS code separately. However, the applicability of this method is limited: (1) It is only specified for CSS code, rather than general qLDPC codes which may contain both X and Z components in one stabilizer check. (2) Although it gives optimal scheduling for X and Z stabilizer checks separately, the compiled circuit is about $2\times$ deeper than the optimal solution. For example, an X-Z separated scheduling of the surface code leads to a depth-8 circuit, which is $2\times$ compared with the best manually designed depth-4 circuit.

Exploiting the sparsity and structural regularity of qLDPC codes, we show that the general constraints of syndrome extraction circuits can be readily expressed as a satisfiability modulo theories (SMT) problem and solved efficiently using modern SMT solvers. Based on this observation, we propose ASC for compilation and end-to-end evaluation of arbitrary qLDPC codes. In summary, our key contributions are:

\begin{itemize}
    \item We analyze and derive a set of general rules for the syndrome extraction circuits of arbitrary qLDPC codes, providing the theoretical foundation of our compiler.
    \item Utilizing the SMT solver, we implement a lightweight yet efficient compiler based on our general rules. This compiler not only significantly reduces human-design overhead but also, for the first time, resolves an open problem in~\cite{Bravyi_2024_arxiv} by providing a definite answer that no depth-6 circuit exists for their listed BB codes.
    \item We develop a comprehensive compilation and evaluation framework, comprising an SMT-based scheduler, a circuit-level noise simulator, and a decoding pipeline for logical level evaluation. Our evaluation demonstrates that ASC can effectively reduce the depth of syndrome extraction circuits and logical error rates, providing further support for theoretical analysis and experimental demonstration.
\end{itemize}

We note that recent works~\cite{peham2025automated, yin2025qecc} explore more comprehensive fault-tolerant circuit compilation frameworks or adopt more advanced techniques such as ancilla bridge. Our work is developed independently and instead focuses on a more concise framework, which enables future integration of constraints from different fault-tolerance considerations or hardware platforms. To facilitate future research,  we open-source both our implementation and evaluation data at \href{https://github.com/iqubit-org/Auto-Stabilizer-Check}{\emph{https://github.com/iqubit-org/Auto-Stabilizer-Check}}.

\section{Background}
\subsection{Stabilizer Formalism}
The concept of \emph{stabilizer} is fundamental to QEC and a certain class of QECCs---stabilizer codes. Rooted in the mathematical frameworks introduced by Gottesman~\cite{gottesman1997stabilizer}, the stabilizer formalism enables describing and manipulating quantum states and operations via a group-theoretic approach. Herein we briefly introduce the stabilizer formalism:

\begin{definition} [Pauli Group]
    The \emph{Pauli group} $P_n$ is composed of n-fold tensor products of $I, X, Y, Z$, with the global phase of $\pm 1$ or $\pm i$.
\end{definition}

\begin{definition}[Stabilizer Group]
    Let $S \subseteq P_n$ be a subgroup of the $n$-qubit Pauli group that does not contain $-I$.
    The subspace
    \begin{equation}
    V_S := \{ \ket{\psi} \in \mathcal{H}_{2^n} \mid s \ket{\psi} = \ket{\psi}, \ \forall s \in S \}
    \end{equation}
    is stabilized by $S$, and $S$ is referred to as the \emph{stabilizer group}. Accordingly, a stabilizer code $\mathcal{C}$ is defined as the codespace stabilized by a corresponding stabilizer group.
\end{definition}



\begin{definition}[Normalizer]
The \emph{normalizer} of a stabilizer group $S \subseteq P_n$ is defined as
\begin{equation}
N(S) = \{ n \in P_n \mid ns = sn,\ \forall s \in S \}.
\end{equation}
\end{definition}
Now we can consider all Pauli errors $E \in P_n$ occurring on stabilizer codes. An error $ E $ is detectable if it anti-commutes with some $s \in S$ ($Es = -sE$), thus producing a non-zero syndrome measurement outcome. On the other hand, if $E \in N(S) \setminus S$, $E$ commutes with all $ s \in S $ and acts as a logical operator on the codespace, making it undetectable by any syndrome measurement. The code distance $d$ is therefore the minimum weight of such operators in $N(S) \setminus S$, ensuring errors of weight less than $\lfloor d/2 \rfloor$ can be corrected.

\subsection{Quantum Low-Density Parity-Check Codes}
We use the notation $[[n, k, d]]$ to denote the physical qubit number $n$, logical qubit number $k$ and code distance $d$ of a stabilizer code. A stabilizer code $\mathcal{C}$ has $n - k$ independent generators $\{g_i\}_{n-k}$, also known as \emph{stabilizer checks}. For example, the $[[7,1,3]]$ Steane code~\cite{steane1996multiple} has the stabilizer generators in \Cref{table:steane code}, and can be represented as the \emph{Tanner Graph}~\cite{breuckmann2021quantum}.

\begin{table}[tbp]
\centering
\caption{Stabilizer generators of the Steane code}
\label{table:steane code}
    \begin{tabular}{c|ccccccc}
        \toprule
        \( g_1 \) & X & X & X & X & I & I & I \\
        \( g_2 \) & X & X & I & I & X & X & I \\
        \( g_3 \) & X & I & X & I & X & I & X \\
        \( g_4 \) & Z & Z & Z & Z & I & I & I \\
        \( g_5 \) & Z & Z & I & I & Z & Z & I \\
        \( g_6 \) & Z & I & Z & I & Z & I & Z \\
        \bottomrule
    \end{tabular}
\end{table}

This representation naturally leads to the \emph{parity check matrix} formulation of the Steane code 
\begin{equation}
H_X = H_Z =
\begin{bmatrix}
1 & 1 & 1 & 1 & 0 & 0 & 0 \\
1 & 1 & 0 & 0 & 1 & 1 & 0 \\
1 & 0 & 1 & 0 & 1 & 0 & 1
\end{bmatrix}
\label{eq: Hx and Hz of Steane Code}
\end{equation}
where an entry of $ 1 $ indicates a non-identity Pauli operator in $\{g_i\}$. Since the Steane code is a CSS code, its stabilizer check can be specified by two independent parity-check matrices $H_X$ and $H_Z$. More generally, a stabilizer code can be represented in the \emph{Binary Symplectic Form} (BSF) as a single matrix $H$ that encodes X, Z or X-Z crossed stabilizers. The shape of $H$ is $(m, 2n)$, denoting $m$ stabilizer checks on $n$ data qubits, where $2n$ is used to represent both X and Z components. 


qLDPC codes are also a family of stabilizer codes, with the weight of \emph{stabilizer checks} bounded by a constant~\cite{breuckmann2021quantum}. In other words, the Hamming weight of each row and each column in $H$ is bounded by a constant factor, corresponding to the maximum degree $\Delta(G)$ of the \emph{Tanner Graph} $G$.

\subsection{Syndrome extraction circuit}
\begin{figure}[tbp]
    \centering
    \resizebox{0.9\columnwidth}{!}{
    \begin{quantikz}[row sep=0.2cm, column sep=0.1cm, align equals at=3]
        \lstick{$q_0$} & \qw & \gate{X} & \qw & \qw & \qw\\
        \lstick{$q_0$} & \qw & \gate{Z} & \qw & \qw & \qw\\
        \lstick{$q_0$} & \qw & \gate{X} & \qw & \qw & \qw\\
        \lstick{$q_0$} & \qw & \gate{Z} & \qw & \qw & \qw\\
        \lstick{$\ket{0}$} & \gate{H} & \ctrl{-4} & \gate{H} & \meter{} & \setwiretype{c}
    \end{quantikz} $=$ \begin{quantikz}[row sep=0.2cm, column sep=0.1cm, align equals at=3]
        \lstick{$q_0$} & \qw & \gate{X} & \qw & \qw & \qw & \qw & \qw & \qw \\
        \lstick{$q_0$} & \qw & \qw & \gate{Z} & \qw & \qw & \qw & \qw & \qw \\
        \lstick{$q_0$} & \qw & \qw & \qw & \gate{X} & \qw & \qw & \qw & \qw \\
        \lstick{$q_0$} & \qw & \qw & \qw & \qw & \gate{Z} & \qw & \qw & \qw \\
        \lstick{$\ket{0}$} & \gate{H} & \ctrl{-4} & \ctrl{-3} & \ctrl{-2} & \ctrl{-1} & \gate{H} & \meter{} & \setwiretype{c}
    \end{quantikz}
    }
    \caption{Illustration of Hadamard test and stabilizer measurement of a given stabilizer generator $g = XZXZ$ as an example.}
    \label{fig:stabilizer check circuits}
\end{figure}
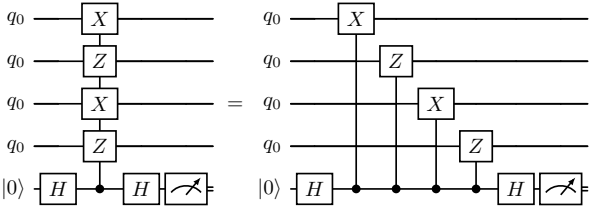

Given a stabilizer check $g \in P_n$ and an $n$-qubit state $\ket{\Psi}$, syndrome extraction refers to the process of measuring the eigenvalue of $g$ on $\ket{\Psi}$. A standard stabilizer check is performed by the syndrome extraction circuit, also known as a syndrome measurement circuit~\cite{bravyi2024high}, which is driven from a standard Hadamard test circuit in~\cite{nielsen2010quantum}
\begin{center}
    \footnotesize
    \begin{quantikz}[row sep=0.2cm, column sep=0.2cm]
         \lstick{data $\ket{\Psi}$} & \qw & \gate{U} & \qw & \qw & \qw \\
         \lstick{ancilla $\ket{0}$} & \gate{H} & \ctrl{-1} & \gate{H} & \meter{} & \setwiretype{c}
    \end{quantikz}
\end{center}
that entangles ancilla and data qubits through $\mathrm{CX}$ or $\mathrm{CZ}$ gates to measure the eigenvalue of $g$ on $\ket{\Psi}$, as depicted in \Cref{fig:stabilizer check circuits}. The syndrome extraction outcomes from ancilla qubits are then decoded to determine the subsequent logical feedback~\cite{litinski2019game}, constituting one of the key components of fault-tolerant quantum computing (FTQC). Since quantum systems are inherently subject to noise---including decoherence, gate imperfections, and environmental interference---that accumulates over the course of circuit execution, longer syndrome extraction circuits significantly increase the risk of introducing additional, uncorrectable errors before error syndromes can be accurately measured. Therefore, optimizing syndrome extraction circuits is crucial for QEC.

\section{General Rules of Syndrome Extraction Circuit for qLDPC Codes}
\label{sec:general_rules}

Consistent with~\cite{bravyi2024high, Bravyi_2024_arxiv}, we define circuit depth as the number of non-overlapping $\mathrm{CX}$ ($\mathrm{CZ}$) layers. We exclude single-qubit gates as they are significantly easier to implement and typically introduce negligible error compared to two-qubit gates. This definition motivates the following questions:
\begin{enumerate}
    \item \emph{What is the minimum depth required for the syndrome extraction circuit of a general qLDPC code?}
    \item \emph{How can we optimize the circuit depth for the syndrome extraction of a general qLDPC code?}
\end{enumerate}

\subsection{General rules}
There are two rules for a syndrome extraction circuit:
\begin{itemize}
    \item Each qubit can participate in at most one two-qubit gate per TICK, where a TICK represents the minimum time step in our circuit. For example, the stabilizer check $XZXZ$ in \Cref{fig:stabilizer check circuits} requires at least 4 TICKs to execute.
    \item The stabilizer check process must align with $H$ to ensure correctness of the QEC process. We will show that the $\mathrm{CX}$/$\mathrm{CZ}$ gates cannot be performed in an arbitrary order.
\end{itemize}

\subsection{Lower bound}
\label{sec:lower_upper_bound}
It is obvious that the lower bound of the syndrome extraction circuit equals the maximum degree of the corresponding \emph{Tanner Graph}, namely $\Delta(G)$. Taking the Steane code as an example, $\Delta(G) = 6$, as data qubit $q_0$ connects 3 X checks and 3 Z checks, each edge represents a two-qubit gate between $q_0$ and one ancilla qubit, thus the minimum number of TICKs to execute all two-qubit gates is at least 6. 

\subsection{X-Z commutativity}
As we have mentioned before, an X-Z separated scheduling is far from optimal. Examining the depth-4 syndrome extraction circuit of the surface code~\cite{Acharya2022SuppressingQE}, we find that there is ample room for optimization. We plot some examples in \Cref{fig:X-Z commutativity}, where $a_1$ and $a_2$ denote two ancilla qubits, corresponding to $XXXX$ and $ZZZZ$ stabilizer checks, respectively. It turns out that swapping the order of $\mathrm{CX}$ and $\mathrm{CZ}$ twice results in a circuit (\Cref{fig:X-Z commutativity} (b)) that is equivalent to the circuit in \Cref{fig:X-Z commutativity} (a). This equivalence allows more $\mathrm{CX}$ and $\mathrm{CZ}$ gates to be executed in parallel. In contrast, swapping their order only once yields a circuit (\Cref{fig:X-Z commutativity} (c)) that is inequivalent to the circuit in \Cref{fig:X-Z commutativity} (a), owing to their anti-commutativity. This observation motivates the following proposition:

\begin{figure}[tbp]
    \centering
    \subfigure[Stabilizer measurement for $X^{\otimes 4}$ and $Z^{\otimes 4}$]{
    \resizebox{0.6\columnwidth}{!}{
        \begin{quantikz}[row sep=0.15cm, column sep=0.4cm, align equals at=3]
        \lstick{$q_0$} & \qw & \gate{X} & \gate{Z} & \qw & \qw & \qw \\
        \lstick{$q_1$} & \qw & \gate{X} & \gate{Z} & \qw & \qw & \qw \\
        \lstick{$q_2$} & \qw & \gate{X} & \gate{Z} & \qw & \qw & \qw \\
        \lstick{$q_3$} & \qw & \gate{X} & \gate{Z} & \qw & \qw & \qw \\
        \lstick{$a_1$: $\ket{0}$} & \gate{H} & \ctrl{-4} & \qw & \gate{H} & \meter{} & \setwiretype{c}\\
        \lstick{$a_2$: $\ket{0}$} & \gate{H} & \qw & \ctrl{-5} & \gate{H} & \meter{} & \setwiretype{c}
        \end{quantikz}
    }
    }
    
    \subfigure[Equivalent circuit]{
    \resizebox{0.49\columnwidth}{!}{
        \begin{quantikz}[row sep=0.15cm, column sep=0.25cm, align equals at=3]
            \lstick{$q_0$} & \qw & \gate{X} & \gate{Z} & \qw & \qw & \qw \\
            \lstick{$q_1$} & \qw & \gate{X} & \gate{Z} & \qw & \qw & \qw \\
            \lstick{$q_2$} & \qw & \qw & \gate{Z} & \gate{X} & \qw & \qw\\
            \lstick{$q_3$} & \qw & \qw & \gate{Z} & \gate{X} & \qw & \qw \\
            \lstick{$a_1$} & \gate{H} & \ctrl{-4} & \qw & \ctrl{-2} & \gate{H} & \qw \\
            \lstick{$a_2$} & \gate{H} & \qw & \ctrl{-5} & \qw & \gate{H} & \qw
        \end{quantikz}
    }
    }\subfigure[Non-equivalent circuit]{
    \resizebox{0.49\columnwidth}{!}{
        \begin{quantikz}[row sep=0.15cm, column sep=0.25cm, align equals at=3]
            \lstick{$q_0$} & \qw & \gate{X} & \gate{Z} & \qw & \qw & \qw \\
            \lstick{$q_1$} & \qw & \gate{X} & \gate{Z} & \qw & \qw & \qw \\
            \lstick{$q_2$} & \qw & \gate{X} & \gate{Z} & \qw & \qw & \qw \\
            \lstick{$q_3$} & \qw & \qw & \gate{Z} & \gate{X} & \qw & \qw\\
            \lstick{$a_1$} & \gate{H} & \ctrl{-4} & \qw & \ctrl{-1} & \gate{H} & \qw\\
            \lstick{$a_2$} & \gate{H} & \qw & \ctrl{-5} & \qw & \gate{H} & \qw
        \end{quantikz}
    }
    }
    \caption{Examples of equivalent and non-equivalent stabilizer measurement circuits for stabilizer checks $g_1 = XXXX$ and $g_2 = ZZZZ$.}
    \label{fig:X-Z commutativity}
\end{figure}
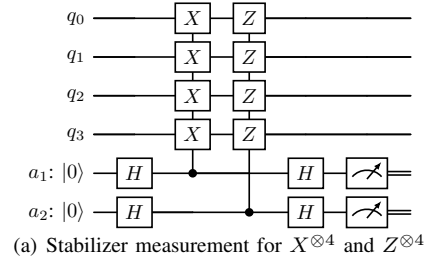
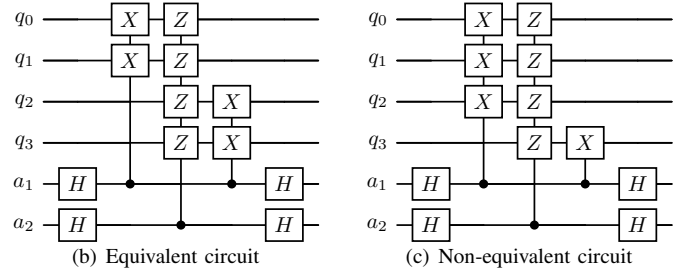

\begin{proposition}[X-Z commutativity]
\label{commutativity-proposition}
For any pair of stabilizer checks, if the parity of the inversion number in the $\mathrm{CX}$-$\mathrm{CZ}$ order across their shared data qubits is even, then all corresponding syndrome extraction circuits are equivalent.
\end{proposition}

We provide a concise proof sketch here. Suppose $\ket{\Phi} = \sum_{x=0}^{2^n-1} c_x \ket{x}$ is an n-qubit stabilizer state on data qubits $\{q_1 \dots q_n\}$ and $\sum_{x=0}^{2^n - 1} \lvert c_x \rvert^2 = 1$, where $c_x \in \mathbb{C}$ are the complex amplitudes associated with each basis state $\ket{x}$. Considering two stabilizers checked by ancilla qubits $a_1, a_2$, the joint state of these $n + 2$ qubits can be written as:
\begin{equation}
    \ket{\Phi} \ket{+} \ket{+} = \frac{1}{2} \sum_{x=0}^{2^n-1} \sum_{y_1=0}^{1} \sum_{y_2=0}^{1} c_x \ket{{x\,y_1\,y_2}}
    \label{eq: joint state}
\end{equation}
where $xy_1y_2$ denotes the concatenation of an $n$-bit string $x = x_1 \dots x_n$, 1-bit strings $y_1$ and $y_2$, thus each basis state in~\eqref{eq: joint state} can be viewed as $\ket{{x\,y_1\,y_2}} = \ket{{x_1\dots x_n\,y_1\,y_2}}$.
Without loss of generality, suppose there exists a $\mathrm{CX}$ gate between $a_1$ and $q_k$, a $\mathrm{CZ}$ gate between $a_2$ and $q_k$. Consider exchanging the order of this $\mathrm{CX}_{a_1, q_k}$ and $\mathrm{CZ}_{a_2, q_k}$, which share the data qubit $q_k$:
\begin{itemize}
    \item If $\mathrm{CX}$ operates ahead of $\mathrm{CZ}$:
    \begin{equation}
    \begin{aligned}
        & \ket{{x_1\dots y_2}}
    \xrightarrow{\mathrm{CX}_{a_1,q_{k}}}
    \ket{{x_1\dots (x_{k}\oplus y_1)\dots y_2}} \\
        & \xrightarrow{\mathrm{CZ}_{a_2,q_{k}}} (-1)^{y_2(x_{k}\oplus y_1)} \ket{{x_1\cdots (x_{k}\oplus y_1)\cdots y_2}}
        \end{aligned}
        \label{eq: CX CZ}
    \end{equation}
    \item If $\mathrm{CZ}$ operates ahead of $\mathrm{CX}$:
    \begin{equation}
        \begin{aligned}
        & \ket{{x_1\dots y_2}}
        \xrightarrow{\mathrm{CZ}_{a_2,q_{k}}}
        (-1)^{y_2 x_{k}} \ket{{x_1\dots y_2}} \\
        &  \xrightarrow{\mathrm{CX}_{a_1,q_{k}}} (-1)^{y_2 x_{k}}\ket{{x_1\cdots (x_{k}\oplus y_1)\cdots y_2}}
        \end{aligned}
        \label{eq: CZ CX}
    \end{equation}
\end{itemize}
Note that~\eqref{eq: CX CZ} and~\eqref{eq: CZ CX} differ only by $(-1)^{y_1 y_2}$. Therefore, an even number of $\mathrm{CZ}$-$\mathrm{CX}$ order exchanges will cancel out the accumulated phase $(-1)^{y_1 y_2}$, yielding a state equivalent to that obtained from separate stabilizer measurements by $a_1$ and $a_2$. Crucially, \Cref{commutativity-proposition} is not limited to the CSS code and applies generally to any type of stabilizer checks, i.e., those involving mixed-X/Z stabilizer generators.

\subsection{Formulated constraints}
We formulate the fundamental constraints of syndrome extraction circuits as follows:

\textbf{Constraint 1.} At each TICK, a qubit may participate in at most one two-qubit gate $\implies$ unique occupation.

\textbf{Constraint 2.} For each $H[i][j] = 1$, there must exist exactly one corresponding two-qubit gate $\implies$ integrality of stabilizer check.

\textbf{Constraint 3.} For any stabilizer check pair, the parity of the inversion number in the $\mathrm{CZ}$-$\mathrm{CX}$ order over their shared data qubits should be even $\implies$ correctness of commutativity.

Note that these constraints serve as general rules for constructing and scheduling syndrome measurement circuits of qLDPC codes, not limited to CSS codes. At this point, we have established concise and complete theoretical guidance for our compiler.

\section{Implementation}

Leveraging the constraints formulated in Section~\ref{sec:general_rules}, we propose ASC's framework, which enables global scheduling of all types of stabilizer checks. For most near-term qLDPC codes, the sparsity of the parity-check matrix $H$ allows it to be efficiently handled by modern SMT solvers. The workflow of ASC is illustrated in~\Cref{fig:workflow}, comprising an SMT-based scheduler in the central processing module, a circuit-level noise simulator for evaluation, and an automated circuit filter.

\subsection{SMT-based scheduler}

\subsubsection{Definition of variables}
The SMT problem formulation of our scheduling task is built upon the following variables:

(\textbf{Boolean Variable}) $H$ has the shape of $(m, 2n)$: the parity-check matrix of the qLDPC code. $H[i][j] = 1 \lor  H[i][j+n]=1$ represents that the $i$-th ancilla qubit has to interact with the $j$-th data qubit at some TICK and only once. $m$ is the number of ancilla qubits (number of stabilizer generators); $n$ is the number of data qubits.

(\textbf{Boolean Variable}) $A$ has the shape of $(m, n, T_{max})$: $A[i][j][k] = 1$ if the $i$-th ancilla qubit has to interact with the $j$-th data qubit at TICK $k$. $T_{max}$ will be defined later.

(\textbf{Integer Variable}) $T$ has the shape of $(m, n)$: the TICK of each $\mathrm{CX}/\mathrm{CZ}$ gate. If ancilla qubit $i$ visits data qubit $j$ at TICK $k$, then $T[i][j] = k \iff A[i][j][k] = 1$. 

Owing to the low-density parity-check property, $H$ is an extremely sparse matrix. To further denote the empty positions $(i, j)$ where no $\mathrm{CX}/\mathrm{CZ}$ gate is required (namely the zeros in $H$) and avoid the scheduling overhead in advance, we just set $T[i][j] = -1$ for all $H[i][j] = 0 \land H[i][j + n] = 0$.

\subsubsection{Optimization Goal}
Our goal is to minimize the circuit depth, i.e., the maximum element in $T$:
\begin{equation}
   \min \; \max_{i,j} \; T[i][j]
   \label{formula:opt_goal}
\end{equation}

\subsubsection{Problem Constraints}
Through this representation, a legal syndrome extraction circuit scheduling must satisfy the following constraints:

\ding{172} Unique occupation of data qubits $j$ at TICK $k$:
\begin{equation}
    \forall j, k \quad \sum_{i} A[i][j][k] \leq 1
    \label{formula:data_qubits_occupation}
\end{equation}

\ding{173} Unique occupation of ancilla qubits $i$ at TICK $k$:
\begin{equation}
    \forall i, k \quad \sum_{j} A[i][j][k] \leq 1
    \label{formula:ancilla_occupation}
\end{equation}

\ding{174} Integrality of stabilizer check:
\begin{equation}
    \forall i,j \quad \sum_{k} A[i][j][k] = H[i][j] \lor H[i][j + n]
    \label{formula:H_complete}
\end{equation}

\ding{175} Correctness of commutativity for all $\{g_i, g_j\}$:
\begin{equation}
\begin{aligned}
    & \forall i, j \ 
    \sum_{l} \mathbb{I}(T[i][l] < T[j][l]) \bmod 2 = 0, 
    l \in \{l \mid \\
    & (H[i][l] \land H[j][l + n]) \oplus (H[i][l + n] \land H[j][l]) = 1\}
\end{aligned}
\label{formula:commutativity}
\end{equation}
where $\mathbb{I}(\cdot)$ equals 1 when the condition holds and 0 otherwise. Equations \eqref{formula:data_qubits_occupation}-\eqref{formula:ancilla_occupation}, \eqref{formula:H_complete} and \eqref{formula:commutativity} completely enforce \textbf{Constraints 1}, \textbf{2} and \textbf{3}, respectively.
\begin{figure}[tbp]
    \centering
    \includegraphics[width=\linewidth]{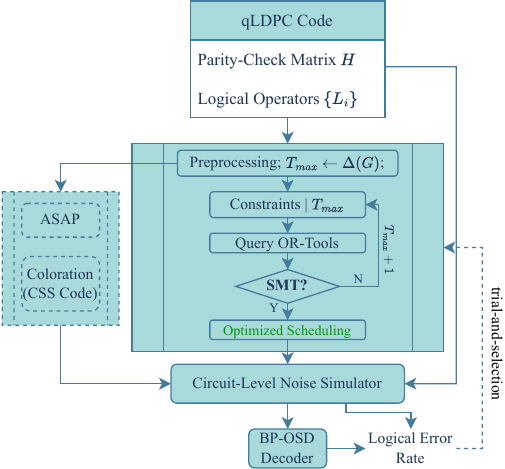}
    \caption{Workflow of ASC with the evaluation pipeline.}
    \label{fig:workflow}
\end{figure}

Instead of directly optimizing the circuit depth in \eqref{formula:opt_goal}, which is usually inefficient, we add one more constraint to confine the search space:

\ding{176} Bound of all TICKs:
\begin{equation}
    \forall i,j \quad T[i][j] \leq T_{max}
    \label{formula:t_max}
\end{equation}
where $T_{max}$ is a constant to bound all $T[i][j]$. We begin with the lower bound established in Section~\ref{sec:lower_upper_bound}, and then search for satisfiability, similar to the strategy in~\cite{tan2024sat}. We incrementally increase the depth step by step until we reach the estimated upper bound, i.e., $2\Delta(G)$. Once a satisfying assignment is found, the search loop terminates and returns the optimized schedule, as described in Algorithm~\ref{alg:sat_solver}. 

\begin{algorithm}[tbp]
\caption{Optimal X-Z crossed circuit scheduling}
\label{alg:sat_solver}
\KwIn{qLDPC check matrices in BSF $H$}
\KwOut{$\mathrm{CX}/\mathrm{CZ}$ scheduling list $T$ of each TICK}
\textbf{Initialize:} $T_{max}$; Conditions $C$; SMT Solver $S$ \;

\For {$T_{max} \gets \Delta(G)$ \KwTo $2 \Delta(G)$} {
    \If{$S(C).timeout()$}{
        \textbf{continue}\;
    }
    \If{$S(C) == True$}{
        \KwRet $T$\;
    }
}
\KwRet False\;
\end{algorithm}

\subsection{Circuit-level noise simulator}
To further evaluate the fault-tolerant performance of the circuits generated by the scheduler, we implement a circuit-level noise simulator using Stim~\cite{Gidney2021StimAF}. After scheduling all $\mathrm{CX}$ and $\mathrm{CZ}$ gates, we build up a standard quantum memory experiment circuit following~\cite{Bravyi_2024_arxiv, bravyi2024high}. This circuit includes $d$ rounds of syndrome extraction (where $d$ is the code distance), and a final measurement round of all data qubits. The syndrome outcomes are passed to the BP-OSD decoder~\cite{panteleev2021degenerate, roffe2020decoding} for decoding, which outputs the \emph{predicted observables} for each logical operator. By comparing these predicted observables with the \emph{actual observables} from the Stim simulator, we compute the logical error rates of the memory experiments---values that reflect the fault-tolerant performance of the circuit.

\subsection{Automated circuit filter}
\label{sec:circ_filter}
We also integrate a simple circuit filtering module for selecting the more fault-tolerant circuits automatically: Since the scheduling produced by the SMT solver exhibits a certain degree of randomness, we can run it multiple times and retain the circuit with the lowest logical error rate, as illustrated by the dashed line \emph{trial-and-selection} in \Cref{fig:workflow}. This might be useful for further theoretical or experimental analysis of different circuit variants of a given qLDPC code.

\section{Evaluation}
In this section, we evaluate the effectiveness of ASC over four aspects: (1) its solution to IBM's open problem regarding the BB code; its ability to (2) reduce circuit depth and (3) improve logical accuracy for general qLDPC codes; and (4) its scalability when applied to larger qLDPC codes.

All experiments are conducted on a dual-socket \emph{Intel Xeon Platinum 8358P} CPU, 
featuring 32 cores per socket, 2 threads per core, a base frequency of 2.60\,GHz, 
and a maximum frequency of 3.40\,GHz. For the SMT solver, we use \emph{OR-Tools}'s CP-SAT from Google~\cite{ortools} with default settings, except that the maximum SMT solving time is set to 2 hours. The circuit-level noise model is the same as~\cite{bravyi2024high}, where each operation of the circuit is followed by a depolarizing channel with strength $p$, including the reset, measurement, idling and $\mathrm{CX}/\mathrm{CZ}$ gates. All memory experiments are conducted under $p = 0.001$, which is the current hardware noise level.

\subsection{IBM's open problem}
We first apply ASC to solve one open problem mentioned in~\cite{Bravyi_2024_arxiv}. The IBM team manually constructed depth-7 circuits by leveraging code symmetries and further identified up to 935 equivalent depth-7 variants through exhaustive computational search, yet they remained uncertain whether a depth-6 circuit could exist. Notably, ASC generates depth-7 circuits for all these BB code examples within 2 hours without timing out. To our knowledge, this constitutes the first concrete evidence supporting the conjecture that no depth-6 circuits exist for these BB codes.

\begin{table}[tbp]
        \centering
        \small
        \caption{Depth-optimal compilation of IBM's BB codes~\cite{bravyi2024high}}
        \begin{tabular}{c|c|c|c|c}
            \toprule
            \textbf{BB Code} & \textbf{ASAP} & \textbf{Color} & \textbf{ASC} & \textbf{ASC Time} \\
            \midrule
            $[[72, 12, 6]]$           & 15 & 12 & \textbf{7} & $< 1$ min  \\
            $[[90, 8, 10]]$           & 15 & 12 & \textbf{7} & $< 1$ min  \\
            $[[108, 8, 10]]$          & 16 & 12 & \textbf{7} & $< 5$ min  \\
            $[[144, 12, 12]]$         & 16 & 12 & \textbf{7} & $< 5$ min  \\
            $[[288, 12, 18]]$         & 16 & 12 & \textbf{7} & $< 5$ min  \\
            $[[360, 12, \leq 24]]$    & 16 & 12 & \textbf{7} & $< 10$ min \\
            $[[756, 16, \leq 34]]$    & 16 & 12 & \textbf{7} & $< 2$ hours\\
            \bottomrule
        \end{tabular}
        \label{tab:BB code}
\end{table}

\subsection{Reduction of circuit depth}
To evaluate the capability of ASC in compiling general qLDPC codes, we select a series of representative examples, including 2 BB codes in~\cite{wang2025demonstration}, 2 GB codes in~\cite{panteleev2021degenerate}, 2 HGP codes in~\cite{pecorari2025high}, and 2 Color codes in~\cite{landahl2011fault}. We show the compilation results in \Cref{tab:general qLDPC code}. Compared with the ASAP scheduling and the coloration-based method, ASC always returns an optimal or near-optimal circuit compilation with depth close to the lower bound $\Delta(G)$, which demonstrates the effectiveness of ASC in searching for optimal syndrome extraction circuits for arbitrary qLDPC codes automatically.

\begin{table*}[tbp]
        \centering
        \small
        \caption{Depth-optimal compilation of some general qLDPC codes}
        \begin{tabular}{c|c|c|c|c|c|c|c|c}
            \toprule
            \textbf{Code} & \textbf{Source} & \textbf{Parameters} & $\Delta(G)$ & \textbf{Manual} &\textbf{ASAP} & \textbf{Color} & \textbf{ASC} & \textbf{ASC Time} \\
            \midrule
            BB Code I  &\cite{wang2025demonstration} & $[[18, 4, 4]]$  & 6 & 7~\cite{wang2025demonstration} & 12 & 12 & \textbf{7} & $< 1$ min  \\
            BB Code II & \cite{wang2025demonstration} & $[[36, 4, 6]]$ & 6 & 7~\cite{wang2025demonstration} & 12 & 12 & \textbf{7} & $< 10$ min  \\
            GB Code I  & \cite{panteleev2021degenerate} & $[[48, 6, 8]]$ & 8 & \emph{unknown} & 21 & 16 & \textbf{9} & ($<$ 2.1 hours) Timeout  \\
            GB Code II & \cite{panteleev2021degenerate} & $[[126, 2, 12]]$ & 4 & \emph{unknown} & 8 & 8 & \textbf{4} & $< 1$ min  \\
            HGP Code I  & \cite{pecorari2025high} & $[[52, 4, 4]]$ & 6 & 10~\cite{pecorari2025high} & 12 & 12 & \textbf{7} & $< 5$ min  \\
            HGP Code II  & \cite{pecorari2025high} & $[[65, 9, 4]]$ & 6 & 10~\cite{pecorari2025high} & 12 & 12 & \textbf{7} & $< 11$ min \\
            Color Code I  & \cite{landahl2011fault} & $[[19, 1, 5]]$ & 6 & \emph{unknown} & 12 & 12 & \textbf{6} & $< 1$ min\\
            Color Code II & \cite{landahl2011fault} & $[[37, 1, 7]]$ & 6 & \emph{unknown} & 12 & 12 & \textbf{6} & $< 1$ min\\
            \bottomrule
        \end{tabular}
        \label{tab:general qLDPC code}
\end{table*}

\subsection{Improvement of logical accuracy}

\begin{figure}[tbp]
    \centering
    \includegraphics[width=\linewidth]{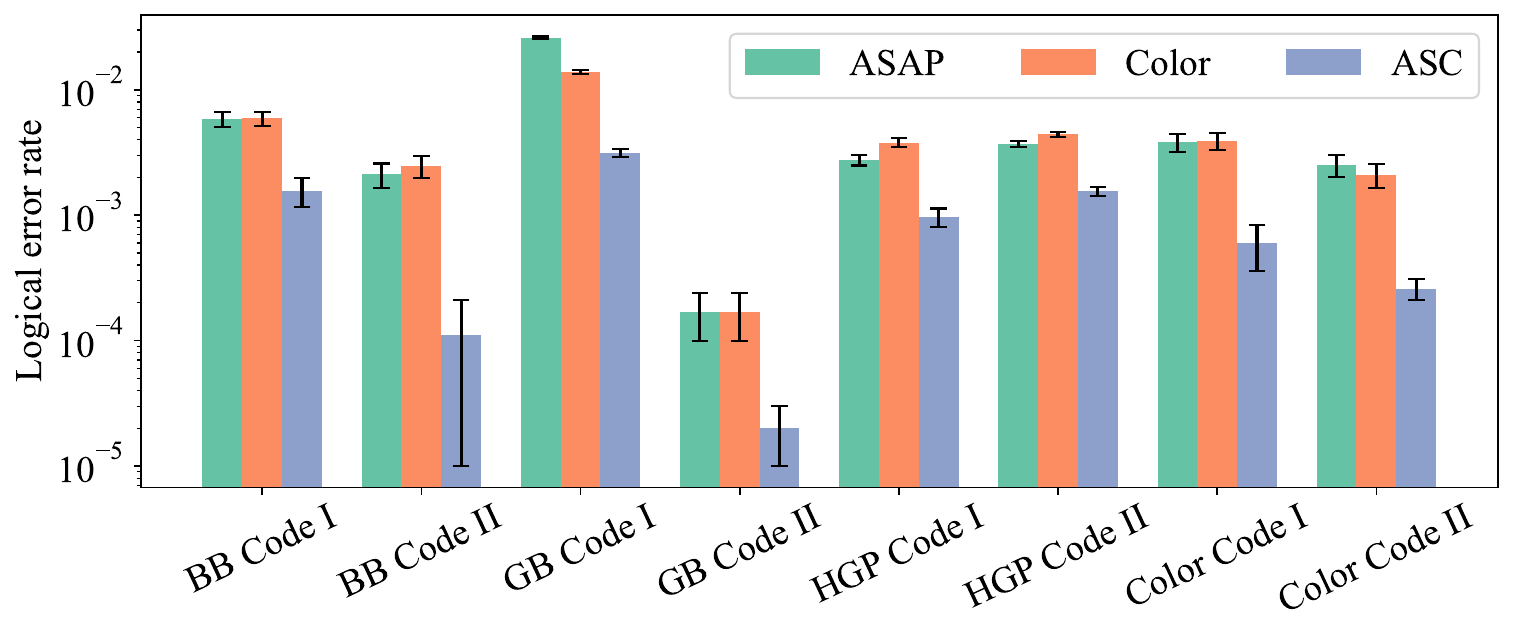}
    \caption{Logical error rate comparison.}
    \label{fig:logical error rate}
\end{figure}

We conduct quantum memory experiments for all qLDPC codes in \Cref{tab:general qLDPC code}. As shown in \Cref{fig:logical error rate}, the logical error rates of circuits from ASC achieve a 7x-8x reduction on average, compared with the ASAP or coloration-based methods. This improvement is primarily attributed to the substantial reduction in idling errors resulting from the nearly halved circuit depth.

\begin{figure}[tbp]
    \centering
    \includegraphics[width=\linewidth]{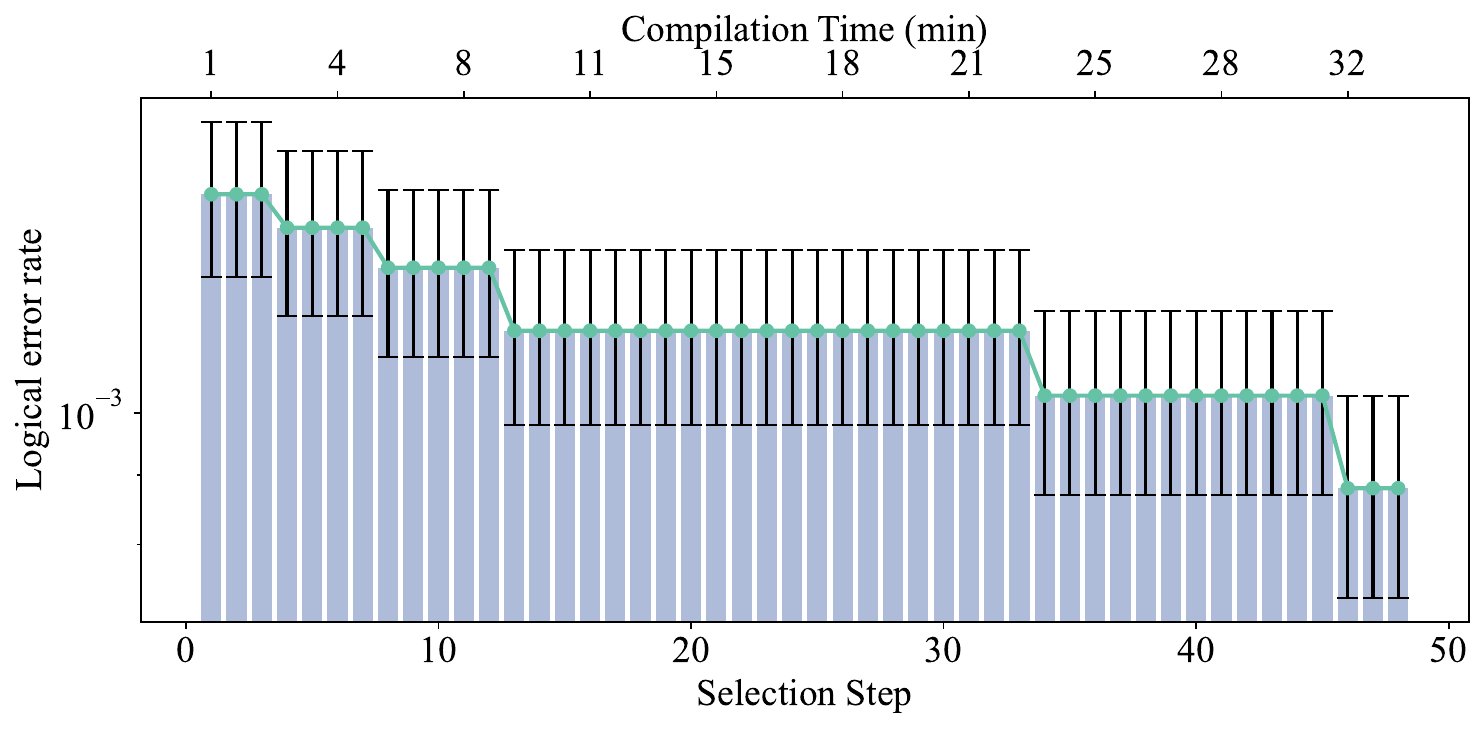}
    \caption{Improvement of fault-tolerance through the automated circuit filter, with a fixed random seed for noise simulation.}
    \label{fig:selection}
\end{figure}

Focusing on the $[[18, 4, 4]]$ code in~\cite{wang2025demonstration}, we also show how to utilize our filter tool described in Section~\ref{sec:circ_filter} to select the circuit with better fault-tolerant performance. As we increase the selection time from 1 to 30 minutes, the corresponding logical error rate of the best selected circuit decreases and gradually converges to optimal, as shown in \Cref{fig:selection}.

\subsection{Scalability}
In \Cref{tab:general qLDPC code}, we find that the compilation time depends not only on the qubit number of a qLDPC code but also on the maximum degree of its \emph{Tanner Graph} (denoted $\Delta(G)$). This is because both factors determine the number of SMT constraints. For example, the $[[48,6,8]]$ GB code uses only 48 data qubits yet times out during compilation, due to its large maximum degree $\Delta(G)=8$. 

\begin{figure}[tbp]
    \centering
    \includegraphics[width=\linewidth]{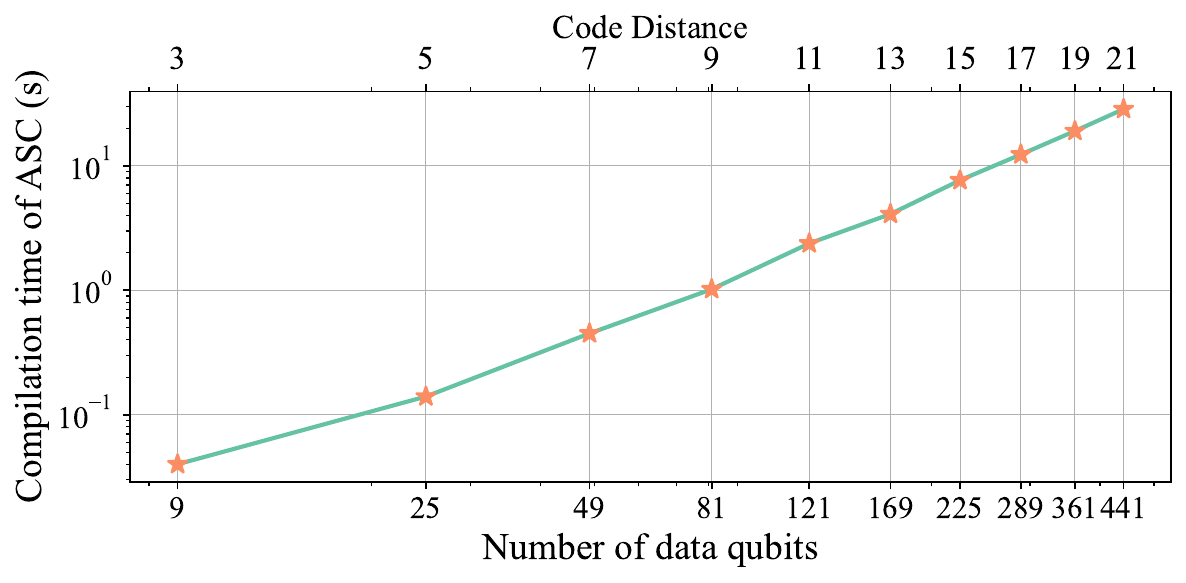}
    \caption{Compilation time scaling of ASC.}
    \label{fig:time scalability}
\end{figure}

We evaluate the compilation scalability of ASC using the surface code: this code family is itself a subset of qLDPC codes, and its members can be characterized by a code distance $d$---all of which admit nearly identical depth-4 optimal syndrome extraction circuits. \Cref{fig:time scalability} illustrates the ASC compilation time against the total number of data qubits $n = d^2$, which scales approximately as a power law yet remains relatively low---staying under one minute even when $d = 21$. This indicates that ASC exploits the sparsity of qLDPC codes with low degree, namely the low density of their parity-check matrices, to achieve efficient SMT solving. However, like other SMT-based approaches, ASC still faces challenges when scaling to larger codes---whether due to more qubits or a larger Tanner graph maximum degree. Nevertheless, compiling large codes could be addressed in future work by combining ASC with localized divide-and-conquer strategies, such as the ancilla bridge method~\cite{chamberland2020topological,wang2024q}.

\section{Conclusion}

In this work, we develop ASC: a lightweight yet efficient compiler and evaluation framework for syndrome extraction circuits in general qLDPC codes, designed around a set of simple rules. For most near-term qLDPC codes, our compiler generates either theoretically optimal or near-optimal solutions within a specified time budget. Via end-to-end evaluations, we demonstrate significant reductions in both circuit depth and logical error rates across a diverse set of qLDPC codes. We further introduce a simple randomized filtering method that automatically produces more fault-tolerant circuits. We anticipate that ASC will serve as a concise, automated qLDPC compilation tool to support both theoretical analysis and experimental demonstrations in quantum error correction.

\section*{Acknowledgment}
The work was supported by National Key Research and Development
Program of China (Grant No.\ 2023YFA1009403), National Natural Science
Foundation of China (Grant No.\ 12347104), Beijing Natural Science Foundation
(Grant No.\ Z220002), the Major Key Project of PCL (Grant No.\ PCL2024A06 and No.\ PCL2025A10), the Shenzhen Science and Technology Program under (Grant No.\ RCJC20231211085918010).

\bibliographystyle{IEEEtran}
\bibliography{reference}

\end{document}